\documentclass[12pt,a4paper]{article}
\usepackage{geometry}
\usepackage[round]{natbib}
\usepackage{authblk}
\usepackage{graphicx}
\usepackage{tabularx}
\usepackage{multirow}
\usepackage{multicol}
\usepackage{amsthm}
\usepackage{algorithm2e}
\usepackage{ulem}
\usepackage{color}
\usepackage{amsmath}
\usepackage{cite}
\usepackage[caption=false]{subfig}

 \usepackage{lipsum}
 \newcommand\blfootnote[1]{
 \begingroup
 \renewcommand\thefootnote{}\footnote{#1}
 \addtocounter{footnote}{-1}
 \endgroup}
\usepackage{mathptmx} 
\usepackage{ragged2e}
\usepackage{sectsty}
\sectionfont{\fontsize{15}{15}\selectfont} 
\subsectionfont{\fontsize{13}{15}\selectfont}
\usepackage{tikz} 
\usetikzlibrary{shapes,arrows}

\usepackage{setspace}

\begin{document}
\begin{titlepage}
\title{\Large Quantitative evaluation of consecutive resilience cycles in stock market performance: A systems-oriented approach}
\centering
\author {\normalsize Junqing Tang\textsuperscript {1*}, Hans R. Heinimann\textsuperscript}
\affil [1] {\scriptsize ETH Zurich, Future Resilient Systems, Singapore-ETH Centre, Singapore, 148602}
\date{}
\end{titlepage}
\maketitle
\begin{abstract}
Financial markets can be seen as complex systems that are constantly evolving and sensitive to external disturbance, such as systemic risks and economic instabilities. Analysis of resilient market performance, therefore, becomes useful for investors. From a systems perspective, this paper proposes a novel function-based resilience metric that considers the effect of two fault-tolerance thresholds: the Robustness Range (RR) and the Elasticity Threshold (ET). We examined the consecutive resilience cycles and their dynamics in the performance of two stock markets, NASDAQ and SSE. The proposed metric was also compared with three well-documented resilience models. The results showed that this new metric could satisfactorily quantify the time-varying resilience cycles in the multi-cycle volatile performance of stock markets while also being more feasible in comparative analysis. Furthermore, analysis of dynamics revealed that those consecutive resilience cycles in market performance were distributed non-linearly, following a power-law behavior in the upper tail. Finally, sensitivity tests demonstrated the large-value resilience cycles were relatively sensitive to changes in RR. In practice, RR could indicate investors' psychological capability to withstand downturns. It supports the observation that perception on the market's resilient responses may vary among investors. This study provides a new tool and valuable insight for researchers, practitioners, and investors when evaluating market performance.

\vspace{1cm}

\raggedright
\text{\textit{Keywords:}}
Resilience; Performance evaluation; Systemic risk; Dynamics and sensitivity; Investor tolerance.
\end{abstract}

\blfootnote{\raggedright \scriptsize{*}{Corresponding author: ETH Zurich, Future Resilient Systems, Singapore-ETH Centre, 1 CREATE Way, CREATE Tower, 138602 Singapore. E-mail: junqing.tang@frs.ethz.ch (J.Tang); hans.heinimann@env.ethz.ch (H.R.Heinimann). Webpage: http://frs.ethz.ch}}

\setlength\parindent{2em}\justify
\vspace{1cm}

\onehalfspacing

\section{Introduction}

The resilience of complex systems has been increasingly studied by policymakers, practitioners, and researchers in recent years. However, resilience is often differently perceived and interpreted in different disciplines~\citep{fisher2015disaster,couzin2018roots}. This has led to the introduction of various frameworks and metrics that have proposed to understand and evaluate this emerging risk-related property via quantitative or qualitative approaches~\citep{bruneau2006overview,linkov2013measurable,nan2014building}.  

Stock markets can be seen as complex evolving systems that constantly undergo uncertain and unexpected disturbance and risks~\citep{kaue2012structure}, causing frequent, volatile, and dynamic market responses. Therefore, performance evaluations concerning factor analysis are not uncommon in stock market research. For instance, the relationship between CEO ownership and stock market performance~\citep{lilienfeld2014ceo} and the relationship between political risks and market equity returns~\citep{lehkonen2015democracy} have been examined. Other factors also include Monetary policy~\citep{gali2015effects}, environmental issues~\citep{joo2017long}, and even the outcome of the US presidential election~\citep{pereira2018trump}. However, evaluating resilient performance in stock markets is missing in these studies.

Apart from those factor-related studies on general market performance, some has focused on the stability and resilience of the markets. For example, ~\citet{leal2017market} studied the stability and resilience of low- and high-frequency trading using agent-based models. They found that regulatory policies can tackle volatility and flash crashes in market performance.~\citet{bookstaber2016toward} studied the impact of decision cycles on overall market resilience and the stochastic features of prices.~\citet{erragragui2018does} investigated the effects of ethics in improving stock market resilience with instability. ~\citet{drakos2011behavioral} studied the diffusion mechanism of terrorist shocks to third countries' stock market responses.

Moreover, there is a research stream in assessing stock market resilience using methods of network science. In this vein, the complex market environment is often represented as networks to study interesting traits, such as market structure and resilience~\citep{kaue2012structure}, individual stock's survivability resilience~\citep{tang2017modeling,tang2018characterisation}, and models for better managing networked markets~\citep{farmer2012complex}. This different vision of stock market analysis offers novel insights into intractable questions, such as explaining why stock markets crash~\citep{sornette2004complex} and the temporal evolutionary process of financial systems~\citep{zhao2018stock}. To our knowledge, the studies of stock market resilience using systems perspectives are mainly focusing on this idea of ``network proxy", systems-oriented approaches still need to be further developed.

In the broader community of economics system studies,~\citet{rose2004defining} has attempted to define several important dimensions of economic resilience to disasters and has established a solid foundation for understanding economic resilience. Furthermore,~\citet{rose2007economic} has proposed a static resilience metric to define system resilience as the ratio of the avoided downturn in overall output and its maximum theoretical downturn. The difficulty in implementing this metric is that the unknown of the estimation of expected post-event system performance~\citep{hosseini2016review}. Even so, this metric has been adapted by others for measuring macroscopic economic resilience~\citep{wein2011economic}. Other excellent and well-acknowledged contributions in the area of economic resilience have also been made by~\citet{duval2007structural,simmie2010economic,martin2011regional}, and~\citet{hill2012economic}. However, most of them are merely qualitative and policy-driven.

On the other hand, in the systems engineering community, the assessment of resilience has emerged with a different viewpoint. To measure system resilience under seismic disturbances,~\citet{bruneau2003framework} have proposed a quantitative metric to assess the ``Resilience-Triangle" in a system's time-series performance~\citep{tierney2007conceptualizing,cimellaro2010framework}. In particular, if one considers that the performance of a system decreases after an external shock but recovers after a certain length of time, then a very straightforward proxy for a system's resilience loss is the decline in performance, which is defined by: (1) failure process (measured from the level of performance immediately before the shock to the lowest performance after the shock), (2) the recovery process (from the lowest performance to the post-event performance after the recovery), and (3) total time between the shock and the end of the post-event recovery. A single successive process of this ``failure and recovery" is also referred to as one ``resilience cycle." This deterministic metric is part of the so-called ``R4" resilience framework, which delineates four functions: Robustness, Redundancy, Resourcefulness, and Rapidity. It has inspired many other performance-based metrics, such as those of ~\citet{ouyang2012time},~\citet{nan2017quantitative} and~\citet{tang2018resilience}. Here, readers are recommended to refer additional review papers for further readings on system resilience, including~\citet{bhamra2011resilience,martin2011resilience,Cere2017}, and~\citet{rus2018resilience}. 

To date, quantitative assessments of resilience can be roughly categorized as general performance-based measures or structure-based models~\citep{hosseini2016review}. Metrics for general measures based on system-level performance are commonly developed as deterministic and probabilistic. This type of metrics usually requires full knowledge of a system's time-series performance profile for subsequent analysis. In this paper, the focus is on deterministic performance-based metrics for measuring consecutive resilience cycles in stock markets. Because stock markets can be viewed as systems, we would like to propose a new approach which adopts a systems-oriented perspective to contribute to the body of knowledge in performance evaluation of stock markets.

There are three evident gaps can be identified from literature review: (1) An innovative coalition that combines systems-based resilience assessment with the evaluation of stock market performance is still needed. (2) Because fault tolerance is an important system attribute for resisting degradation and maintaining performance~\citep{gonzalez1997adaptive}, it should be intuitively incorporated into resilience metrics. However, most well-established metrics have ignored the effects of the tolerance capability, which is a top-level system capability to ``tolerate downturns" during shocks. And (3) the exploration of the dynamic features of consecutive resilience cycles in time-series market performance has been largely overlooked.

To bridge these gaps, this paper proposes a metric based on a comprehensive resilience framework from a systems perspective, which ontologically identifies a series of system-level elemental functions. The consideration of two system's fault-tolerance thresholds, Robustness Range (RR) and Elasticity Threshold (ET), was quantitatively incorporated into the metric, and three existing metrics were selected for comparison purpose. The empirical study was conducted by utilizing the daily closing values of two stock market indexes: NASDAQ and SSE (Shanghai market in China). Data were collected from open sources over a 5-year span, from 2013 to 2018. 

To illustrate the merits of the metrics, the proposed and selected metrics were tested in these two empirical cases. The dynamics of the consecutive resilience cycles and the model sensitivity were demonstrated by analyzing distributions and robustness analysis. Eventually, insights of resilient performance in the markets and the effect of investors' psychological perception on quantified resilience indicators were remarked. 

The main contributions of this paper are summarized as:
\begin{enumerate}

\item The paper proposes a novel and relatively better performance-based metric, from a systems perspective, to quantify consecutive resilient responses in stock markets, which considers the effects of a system's capability of fault tolerance. This bridges one of the gaps in the literature and enriches the toolkit for performance evaluation studies in financial markets.

\item The paper provides insights on the stochastic characters of time-varying resilient behaviors in volatile market performance and draws inferences about market resilience as well as investors' psychological effects. 

\item Because of the generality of this systems-oriented approach, the proposed methodology could have high transferability to other fields of research and contributes to broader scientific understanding about performance-based resilience.
\end{enumerate}

In the upper half of this paper, we construct the resilience metric using a language of systems science in Section 2-4, including terminologies and viewpoints. Evidence related to financial stock markets is given where it is necessary for clarification. In the second half, Section 5-6, we demonstrate the applicability of the proposed metric by empirically studying the market resilience in the selected market cases, and conduct dynamics analysis and sensitivity tests. Finally, we end this paper by summarizing the main findings.


\section{Function-based resilience framework}
\label{framework}
\begin{figure}[htb!]
        \centering
	\includegraphics[width=0.9\textwidth]{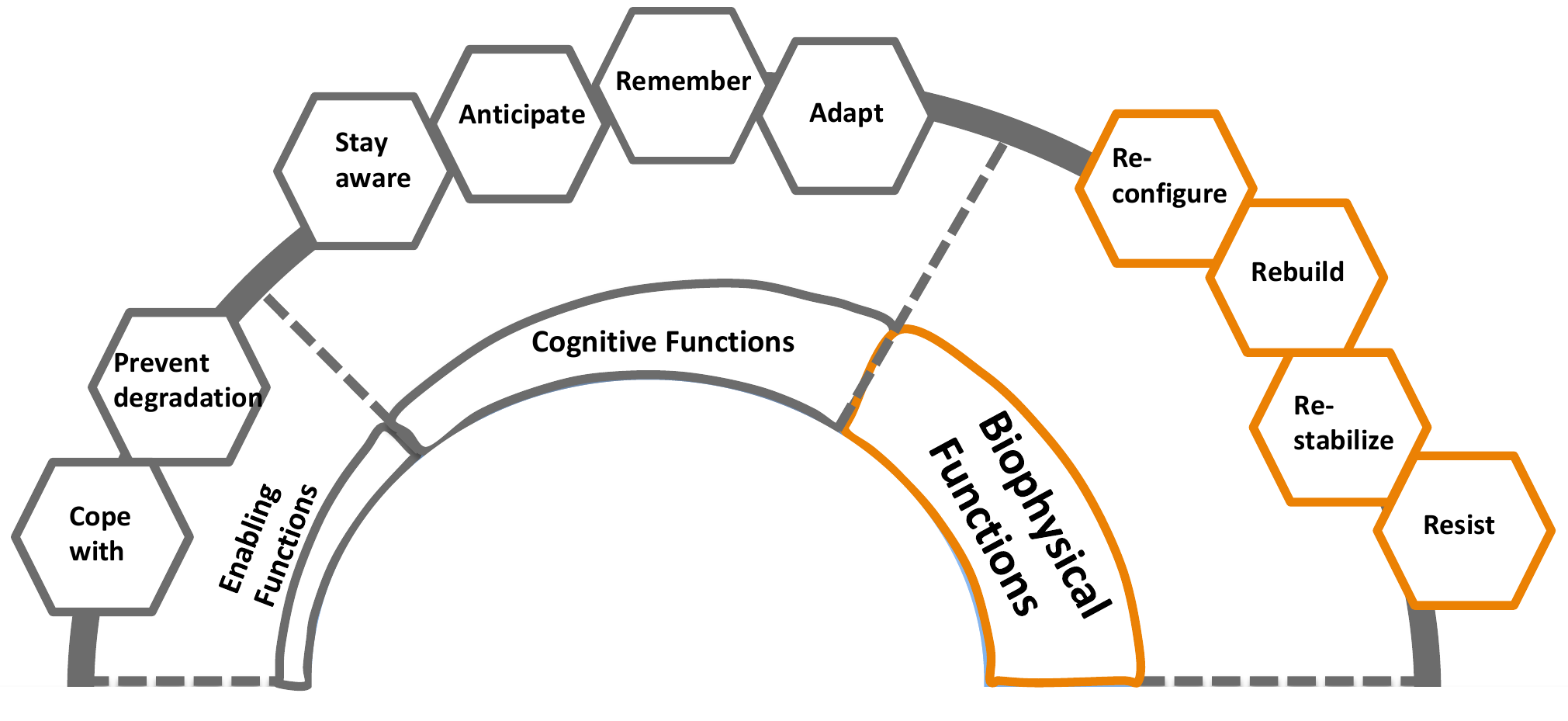}
	\caption{Reconstructed resilience framework with three classes of functions. Adapted from~\citet{HeinimannHatfield}.}
	\label{HansFramework} 
\end{figure}

Recent research by \citet{HeinimannHatfield} has framed and defined resilience in terms of transferable functions by following the ``Structure-System" paradigm of systems engineering and providing a comprehensive multi-dimensional consideration (Fig~\ref{HansFramework}). Let us briefly explain this framework. The biophysical functions of a system's resilient behavior include resistance, re-stabilization of critical performance, and the rebuilding and reconfiguration of that performance. Secondly, from the perspective of a socio-technical system, cognitive resilience must feature awareness, anticipation, remembrance of useful action, learning, and adapting. Finally, resilience can then be incorporated into long-term macroscopic enabling properties, with the goal of coping with the changing environment and preventing degradation in the long run. 

Because the scope of the paper is focusing on performance-based assessments of microscopic resilience cycles and does not involve any cognitive study, we adopted those biophysical functions as the applied framework for pragmatic considerations in general~\citep{HeinimannHatfield}. 

Biophysical functions provide an aggregated characterization of a single ``resilience cycle" (or ``bathtub" shape) in typical system performance. This particular shape can be triggered by a shock (either external or internal), followed by immediate absorption and post-event reaction by the system that includes the abilities to restore and adapt~\citep{biringer2016critical}. The framework assigns these responsive capabilities as broader concepts, e.g., the capability to resist, re-stabilize, rebuild and re-configure, which then questions our understanding of a system's behavior as: 
\begin{itemize}
\item \textit{Resist}: How is the system's ability to resist the negative effect of the shock on its performance? This function acts as an aggregated outcome of a system's overall ability to resist and should include all sources of resisting efforts, such as the ability to be robust, ability to actively prevent disturbance, and ability to maintain stability, etc.

\item \textit{Re-stabilize}: Can the system respond to a shock by stabilizing its performance, and how much effort is needed to regain or maintain this post-event stability? This function describes the system's ability to stabilize the level of functionality so that the adverse effects caused by disturbances would not fatally compromise the system.

\item \textit{Rebuild}: Once the performance is stabilized and total collapse prevented, how can the system rebuild its performance after the shock? This function is essential because there would be minimal room to discuss resilience if the systems do not acquire the ability to rebuild its performance afterward.

\item \textit{Re-configure}: After surviving the shock, how can the system's overall performance be re-organized and updated? In this way, the adaptive behavior would be possible, and the overall system would be more resilient than before. 
\end{itemize}

In this paper, these four biophysical functions were termed as ``elemental functions" (resistance, re-stabilization, rebuilding, and re-configuration) because they are the key characteristics of a system's resilient response during a disruption event. From the perspective of system performance, they crystallize the system states from ``pre-event" via ``during-event" to ``post-event" according to the original framework.\footnote{It is important to note that this framework of resilience, including these elemental functions, may not be universally accepted by different researchers. As described in Section 1, the exact framework of system resilience is a controversial topic. Here, we adopted this framework because of its comprehensiveness and generality.}

\section{Tolerance thresholds}
Primarily described as a generic system character in systems engineering, fault tolerance -- the ability of a system to continue its operation with acceptable performance in an event that some of its performance is compromized~\citep{gonzalez1997adaptive} -- also plays an indispensable role in sculpting system resilience. Here, two thresholds can be defined which contribute to general tolerance capability of a system, Robustness Range (RR) and Elasticity Threshold (ET). These thresholds represent two levels of tolerance.

\textbf{First level:} RR acts as the first level of tolerance, where the compromise is not fatal. As described, this threshold depicts a ``system's ability to maintain its performance within an acceptable range of degradation"~\citep{haimes2009complex}. Therefore, as long as performance is maintained within the RR, it is considered acceptable. In other words, this level of tolerance represents the system's capability to withstand and resist the negative effects of disturbances. More importantly, this concept is generic across stock markets. For example, stock market indexes incorporate an intuitive ``safe range" so that investors can perceive attentive losses or gains that fall outside of this range~\citep{johansen2010shocks,filimonov2015power}\footnote{This could also be true in other systems. For example, water distribution systems have high- and low-pressure boundaries that confine a robustness range in order to maintain acceptable functionality~\citep{hashimoto1982reliability}.}. 

\textbf{Second level:} Similar to the concept of ``Elasticity" in the field of mechanics, a lower threshold, ET, should also be incorporated to denote phase shifts in performance states. This is the second level of fault tolerance which can be seen as the last ``defense line" where the downturn of a system's performance could still be tolerable to some extent. If the performance dropped below this threshold, it is assumed that this phase transition (or de-stabilization) would cost significant efforts and time for a system to regain its performance. For instance, financial stock markets can experience well-documented regime shifts due to unexpected disturbances, e.g., financial crashes~\citep{kaizoji2000speculative,wilinski2013structural}\footnote{Some other social and engineering systems are also known to have similar phase transition phenomena~\citep{kacperski1996phase}, such as transportation systems and social networks.}. 

\begin{figure}[htb!]
	\centering
	\includegraphics[width=1\textwidth]{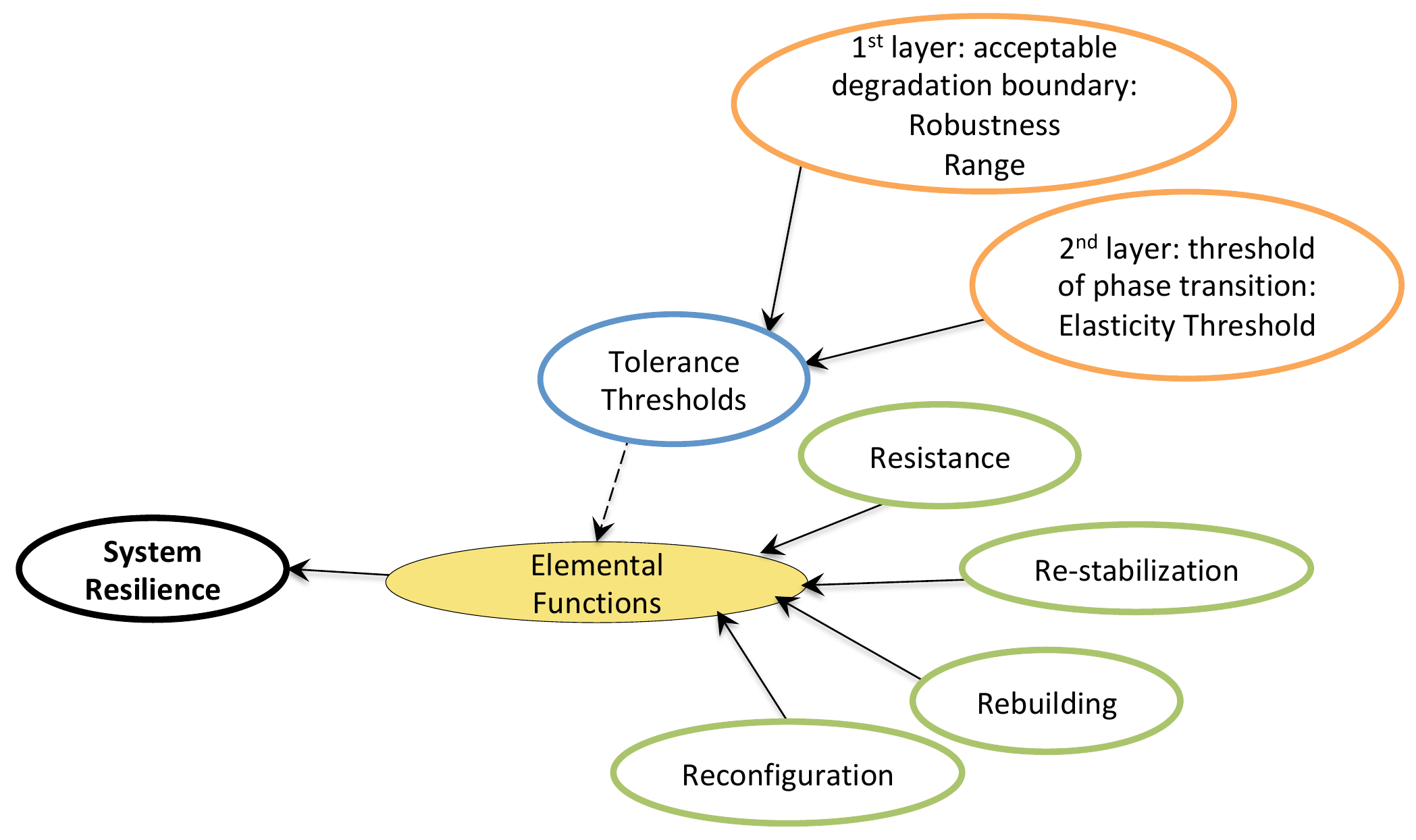}
	\caption{System resilience decomposed as fault tolerance and elemental functions. Notes: black arrows indicate ontological interdependence or "contribute-to" relationships; dotted arrow, interactive effect between tolerance thresholds and elemental functions.}
	\label{Flowchart} 
\end{figure}

Fig~\ref{Flowchart} illustrates an ontological relationship between these two thresholds and the elemental functions established in the previous section. The tolerance thresholds influence the elemental functions of a system and, consequently, its resilience. Therefore, those thresholds should be incorporated into the process of measuring resilient responses.

\section{Metric construction}
\label{resilience attributes}

\begin{figure}[htb!]
	\centering
	\includegraphics[width=1\textwidth]{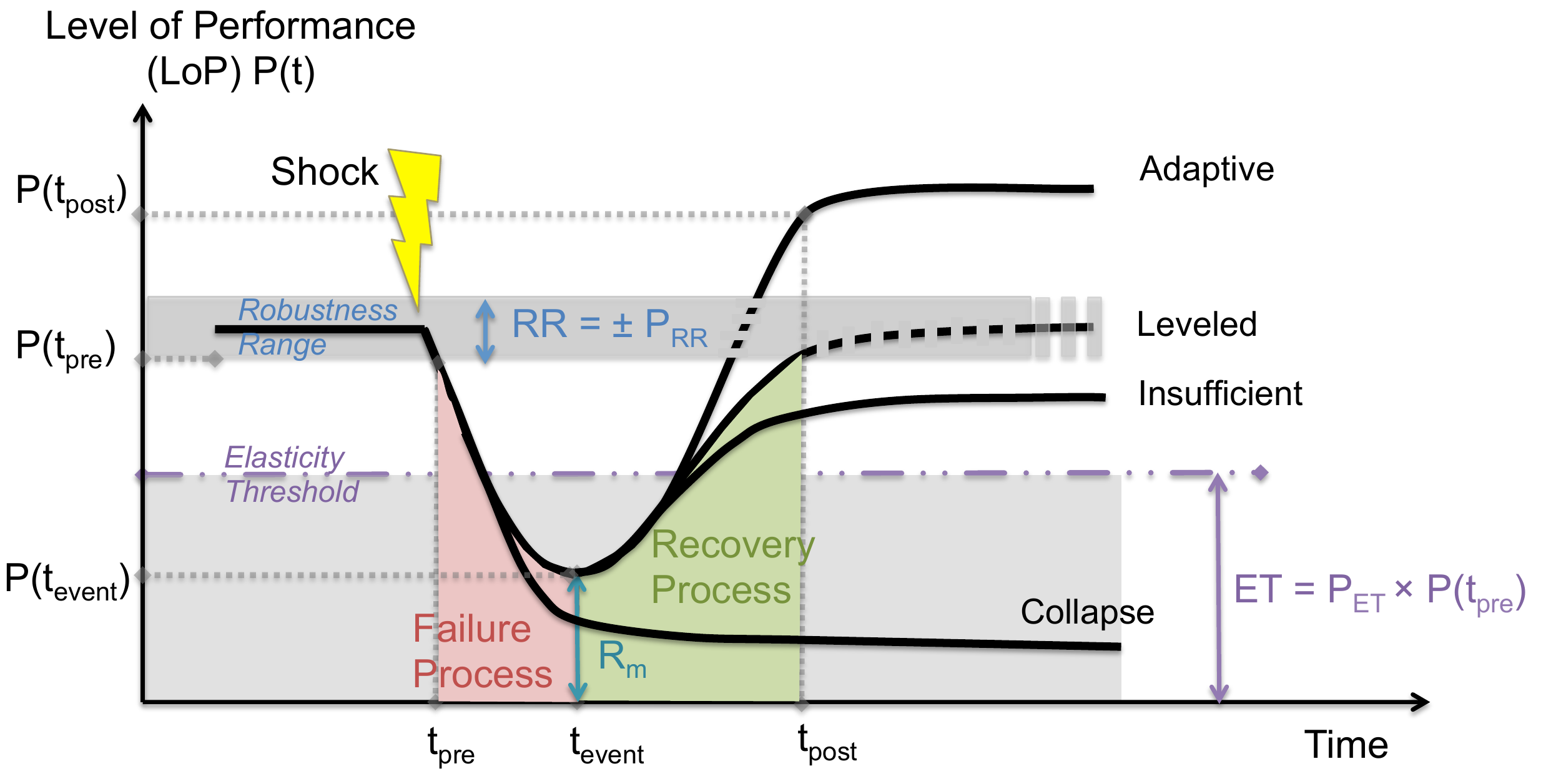}
	\caption{Example of resilience cycle associated with 4 types of post-event performance.}
	\label{scenarios} 
\end{figure}

This section presents the quantification of functions and thresholds to form the proposed metric. Firstly, some general assumptions are necessary: (1) Assume that a shock happens at time $t_{pre}$ with the level of performance (LoP) at $P(t_{pre})$, the point at which the downturn/failure process starts (red area in Fig.~\ref{scenarios}), and the system's LoP begins to decrease. At time $t_{event}$, the LoP would reach its minimum level at $P(t_{event})$, which then initiates the upturn/recovery segment (green area in Fig.~\ref{scenarios}). The recovery would eventually complete at time $t_{post}$ with a post-event performance of $P(t_{post})$; And (2) Discussions on shock features, such as sources, intensities, forms, etc., are not in the scope of this paper, we simply symbolized ``shocks" as a general outcome of stimuli and disturbance that could reflect on the overall performance in a broader sense.

A multi-cycle hybrid performance (MCHP) contains a series of resilience cycles with a mixed combination of different types of post-event performance. The overall system-level performance of a stock market can be represented as a time-series evolving index, which can be seen as a concrete example of MCHP. Fig~\ref{scenarios} presents four possible post-event performance that can be widely observed in a typical resilience cycle~\citep{Heinimann2016}. They can be outlined as follows: 

\begin{itemize}
\item \textit{Collapse}: This type of $P(t_{post})$ is equivalent to ``worst-case recovery" or ``no recovery," which represents a situation in which a system's performance cannot organize any form of effective recovery after the shock. Here, notice that the term ``Collapse" did not necessarily indicate physical damages, but rather served as a representation of ``collapse of LoP"; 

\item \textit{Insufficient}: As is self-evident, this type of $P(t_{post})$ indicates a partial recovery when compared with $P(t_{pre})$, such that $P(t_{pre})$ $>$ $P(t_{post})$; 

\item \textit{Leveled}: This is the most widely studied paradigm in the field of resilience, and is used to describe the case at which performance loss is fully restored at $P(t_{post})$;

\item \textit{Adaptive}: Learning and adapting behaviors are pervasive in some highly evolving (e.g., complex adaptive and intelligent systems) or volatile systems (such as stock markets). Such systems often foster cases in which $P(t_{post})$ exceeds the level of $P(t_{pre})$ to promote growth in overall performance.
\end{itemize}

\subsection{Characterization of tolerance thresholds}
Based on this outline of the possible post-event performance in a resilience cycle, the tolerance thresholds can then be defined as follows:

\emph{Robustness Range ($RR$)}: The wider the RR, the more resilient the performance. During the process of calculations, the range is computed as a percentage, $P_{RR}$, and is expressed as:
\begin{equation}
RR = \pm P_{RR}
\end{equation}

\emph{Elasticity Threshold ($ET$)}: If LoP falls below the ET (when the phase transition occurs), then extra efforts may be needed to help the system re-gain its LoP. It can be expressed as a certain percentage, $P_{ET}$, of the $P(t_{pre})$ in a resilience cycle.
\begin{equation}
ET = P_{ET} \times P(t_{pre})
\end{equation}

\subsection{Quantification of elemental functions}
Once the tolerance thresholds were set, the elemental functions can be quantified accordingly as follows:

\emph{Resistance ($R_m$)}: Resistance should be a total measure of the capability to resist the negative impact caused by a disturbance with the capability to maintain the LoP immediately after that occurrence. It is an aggregated outcome of system's overall ability to prevent negative effects of shocks, which could be approximated as the minimum LoP $P(t_{event})$ during the entire process from $t_{pre}$ to $t_{post}$~\citep{nan2017quantitative} plus the maintaining effects contributed from the robustness range. This measure identifies the maximum impact of disruptive events, and use the remaining capacities as a proxy for the system's current ability to resist.
\begin{equation}
R_{m}=P(t_{event}) + RR \quad for \quad t\in[t_{pre}, t_{post}]
\end{equation}

\emph{Re-stabilization ($R_e$)}: This function determines the need to regain stability when a phase transition occurs in performance. Intuitively, if the level drops below ET (assuming that the transition happens when LoP falls below ET), then the need for $R_e$ would be considered large because more stability has been lost after disruptive events. Therefore, a high $R_{e}$ should indicate low resilience and $1-R_{e}$ would denote the remaining available stabilizing capacity. If the minimum point of LoP was above ET, it is assumed that its $R_{e}$ shall be zero. Hence:
\begin{equation}
R_{e}=\frac{ET-P(t_{event})}{P(t_{pre})-P(t_{event})} \quad if \quad ET > min[P(t)]
\end{equation}

\emph{Rebuilding ($R_d$)}: The relative rapidity of the entire recovery process is considered a sensible measure for assessing and quantifying the rebuilding function in a resilience cycle. A quick recovery process means that the system is adequate to re-organize, rearrange, and re-establish. As defined by \citet{nan2017quantitative}, let $S_{f}$ and $S_{r}$ represent the rapidity in downturn and upturn processes, respectively, then the rebuilding capability is measured as:
\begin{equation}
R_{d}=\frac{S_{r}}{S_{f}}
\end{equation}
where, 
\begin{equation}
S_{f}=\frac{P(t_{pre})-P(t_{event})}{t_{event}-t_{pre}} \quad and \quad
S_{r}=\frac{P(t_{post})-P(t_{event})}{t_{post}-t_{event}}
\end{equation}

\emph{Reconfiguration ($R_s$)}: This attribute is sometimes labeled as the ``Recovery scenarios"~\citep{tang2018resilience} or ``Recovery path" in previous studies~\citep{nan2017quantitative}. Nevertheless, this term essentially characterizes various reconfiguration results after a recovery. It can be defined, straightforward, as follows:
\begin{equation}
\label{s}
R_{s}=\frac{P(t_{post})-P(t_{event})}{P(t_{pre})-P(t_{event})}
\end{equation}
where, $P(t_{post})-P(t_{event})$ is the LoP restoration during the recovery process and $P(t_{pre})-P(t_{event})$ is the LoP loss in the failure process.

\subsection{The proposed metric}
After defining and quantifying all of these elemental functions accordingly, we expressed the proposed metric for offering a resilience indicator (RI) in each resilience cycle as: 

\begin{equation}
RI(t)=f(R_{m}, 1-R_{e}, R_{d}, R_{s})=R_{m}\times (1-R_{e})\times R_{d}\times R_{s}
\label{metric}
\end{equation}

The rationale of this function-based metric is explained in several steps. First, a multiplicative form was designed to avoid casual bias because the weight of each function is often unobtainable. Second, resilience should intuitively be proportional to the resistance capability, i.e., a larger $R_{m}$ would denote a larger RI. Third, in the case of recovery per unit time, $S_{r}$, is larger than the loss/failure per unit time $S_{f}$. Therefore, system performance should exhibit resilient behavior due to a rapid rebuilding of the LoP. This would then lead to a larger RI value. Fourth, if minimum performance was below ET, then $(1-R_{e})$ represents the available capacity for stabilizing performance during the phase transition. Finally, $R_{s}$ differentiates various post-event cases where the value for adaptive behavior would be larger than that calculated for other types of recovery, such that the following order occurred: ``Adaptive" $>$ ``Leveled" $>$ ``Insufficient" $>$ ``Collapse."

The metric is supposed to be dimensionless (after normalization), and non-negative whose value will be zero in the following cases:
(1) there is no recovery (i.e., a collapsed case), meaning that both $S_{r}$ and $R_{s}$ are zero and system performance either shows no resilience to cope with a shock or the LoP is stabilized at the lowest point and never re-bounce;
(2) given that the lowest point of LoP reaches zero, i.e., $R_{m}$ and $(1-R_{e})$ are zero. Meanwhile, it may only take a second to realize that this metric could be mathematically correct if and only if the downturn process existed, that is, the $S_{f}$ is not zero, so that $R_{d}$ and $R_{s}$ would each have a non-zero denominator.

\subsection{Selected metrics for comparison}
\label{comparing_metrics}
Three well-established resilience metrics are selected for comparison exercises. The selection criteria were based on their characteristics and popularity.

The first metric, R1, was the ``Resilience-Triangle" metric proposed by \citet{bruneau2003framework} as described in the Section 1. It depicts the total loss of system resilience based on the difference in area between 100\% functionality and the actual performance $Q(t)$. Because it is characterized by the triangular area of performance loss, its value should always be greater than zero. Given that $Q(t)$ is the actual LoP at time $t$, $t_{0}$ is the $t_{pre}$, and $t_{1}$ is the $t_{post}$, then R1 is expressed as:

\begin{equation}
R_{1}=\int_{t_{0}}^{t_{1}} [100-Q(t)]dt
\end{equation}

The second metric, R2, was firstly proposed by \citet{ouyang2012time}. Although its very essence is similar to R1, modifications have been made to the representation of resilience. This metric confines its domain range from zero to one as a ratio between the target (normally 100\% LoP) and the actual LoP. The expectation sign in its original computational form, as shown, is designed to obtain the overall expected value of a series of resilience cycles.

\begin{equation}
R_{2}=E[\frac{\int_{0}^{T} P(t)dt}{\int_{0}^{T} TP(t)dt}]
\end{equation}
where, $T$ is the total time of observation, $P(t)$ is the actual LoP at time $t$, and $TP(t)$ is the target LoP at time $t$.  

The third metric R3 was initially presented by \citet{francis2014metric}. It is not an ``area-based" metric in which one considers the rate of recovery and LoP at critical points are considered. Similar to the proposed metric, it depicts resilience with various post-event performance and is a deterministic metric that can capture adaptive behaviors. 

\begin{equation}
R_{3}=S_{p}\times \frac{F_{r}}{F_{0}}\times \frac{F_{d}}{F_{0}}
\end{equation}
where, $S_{p}$ is the speed of recovery, $F_{r}$ is the LoP after recovery, $F_{0}$ is the LoP before a disaster, and $F_{d}$ is the lowest point of LoP.

\section{Empirical study}
\label{case studies}
This section presents a step-by-step empirical study of the proposed metric in the time-series performance of two stock market indexes, NASDAQ and SSE. First step: data collection, normalization, and de-noizing are introduced. Second step: initial settings, such as identification of consecutive resilience cycles and determination of tolerance thresholds, are presented. Third step: quantification results and comparative analysis are demonstrated, and we comment on the performance of different metrics in both study cases. Finally, we analyze the dynamics of the quantified resilience cycles and explore the metric's robustness in sensitivity tests.

\subsection{Data description and pre-processing}
The empirical data are daily closing values of the NASDAQ Composite Index (INDEXNASDAQ:.IXIC) and SSE Composite Index (SHA: 000001), collected from Yahoo Finance \citep{YahooFinance} with a specific period from 16 September 2013 to 16 April 2018. The time-series performance of a stock market would be an excellent example for multi-cycle cases with hybrid post-event performance.

In general, a common platform for evaluating LoP is a normalized range between zero and one~\citep{nan2017quantitative}. Therefore, the pre-processing step begin by normalizing the market performance index so that the LoP (y-axis) could be confined to the range [0,1]. In this way, it facilitates the comparison between different metrics as well. Here, given a LoP, $P(t_{i})$, at time $t_{i}$ and conducting a simple statistical normalization, the normalized LoP can be expressed as $P(t_{i})/max[P(t)]$. Normalization of the x-axis was done by using a daily interval as the normalized x-axis. 

\begin{figure}[htb!]
	\centering
	\includegraphics[width=1\textwidth]{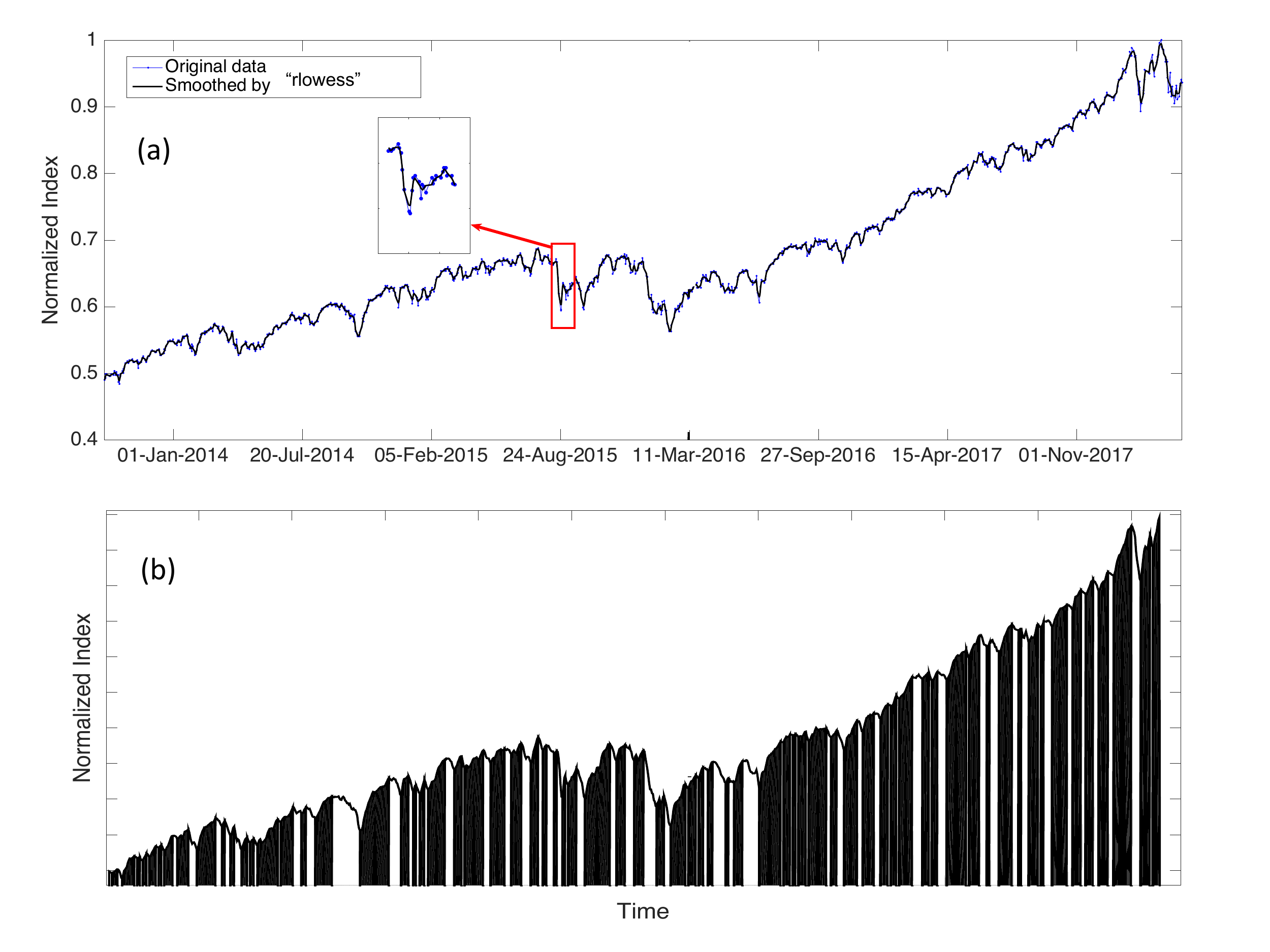}
	\caption{De-noised data and identification of cycles. (a) original data and de-noised result smoothed by ``rlowess" algorithm. (b) identified resilience cycles (black shading, drawups; white shading, drawdowns).}
	\label{data_description} 
\end{figure}

For a volatile performance in a stock market, measurements can be sensitive to background noise~\citep{filimonov2015power}. Therefore, we applied the ``rlowess" algorithm in the Matlab toolbox~\citep{CurveFitting}, a robust version of a local regression using weighted linear least squares and a first-order polynomial model, to de-noise the performance data with a window span of four days. Taking the NASDAQ as an illustrative example, the normalized and de-noised index performance is shown in Fig~\ref{data_description}~(a).

\subsection{Initial settings}
Implementing metrics requires one to identify, in advance, meaningful ``up-and-down" resilience cycles in the market performance~\citep{johansen1998stock,johansen2000large}. To address this, \citet{johansen2010shocks} have developed a straightforward filtering algorithm, called $\tau-$filter. Essentially, if the LoP drops successively for fewer than $\tau$ days, then this drawdown can be neglected as a minor fluctuation. Simply put, this filtering process cleans out meaningless cycles by observing their duration. Here, $\tau$ was set as three days, so that any drawdown or drawup that lasted less than three days would be merged with its neighbor portion. The final identification outcome of resilience cycles in the NASDAQ over five years of observation is presented in Fig~\ref{data_description}~(b) (with black drawups and white drawdowns).

The practical RR and ET for each resilience cycle were initially determined as 0.01\% and 80\%, respectively. As suggested by \citet{filimonov2015power} and \citet{johansen2010shocks}, investors could be insensitive to any return fluctuation of 0.01\% but will probably be hurt, either financially or psychologically, by a decline of 40\%. We assumed that half of this empirical value would cause a phase transition, therefore ET = 1 - 20\% = 80\%.

\subsection{Quantification and comparative analysis}
\begin{figure}[htb!]
	\centering
	\includegraphics[width=1\textwidth]{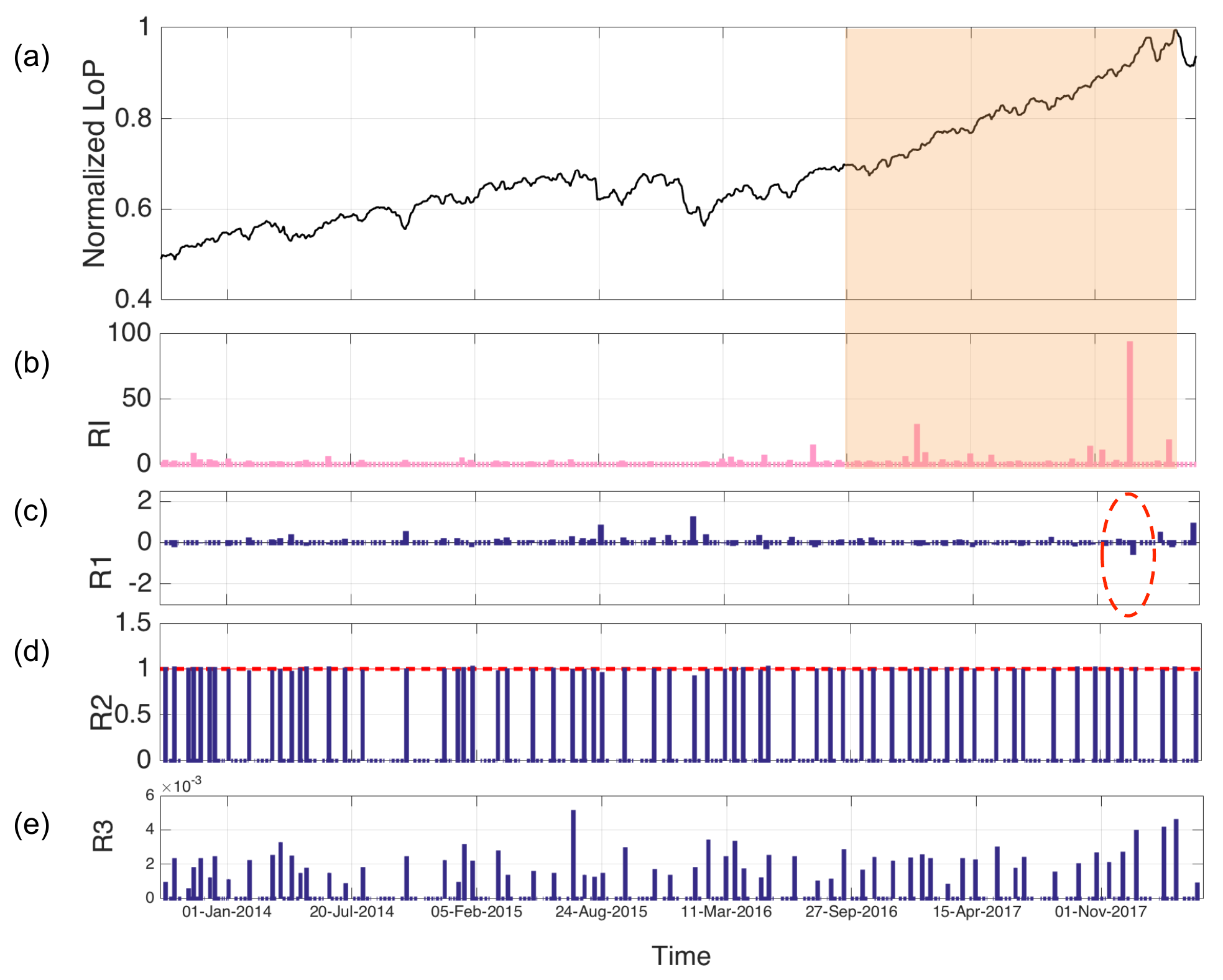}
	\caption{Results from comparisons among four tested metrics, based on the NASDAQ case. (a) normalised LoP, (b) RI, (c) R1, (d) R2, and (e) R3. Obvious defects are highlighted in red. Indicators are plotted at the $t_{event}$ in each cycle.}
	\label{resilience_comparison1} 
\end{figure}

Fig~\ref{resilience_comparison1} presents the quantification outcomes from the NASDAQ case, with subplot~(a) showing normalized performance and (b-e) exhibiting the results from RI, R1, R2, and R3, respectively (readers can refer Appendix A for a detailed illustrative example on how to implement the proposed metric. The implementation procedures of selected metrics are documented in the corresponding literature as cited in the previous section). Upon cross-referencing with the first 10 numerical values in Table~\ref{resilience_comparison_table}, one can see from the figure that RI provided not only appropriate scores but also outperformed other metrics by assigning larger scores to continuous increasing portions (see indicators in the highlighted segment). Most importantly, RI quantified the resilient performance of the market in an intuitively appropriate way. For example, it is obvious that the overall trend of the highlighted segment is more resilient because of NASDAQ's continuous and strong adaptive performance during that period. Whereas this was well-captured by the proposed metric, the others either failed to indicate such time-varying information (Subplot (c) and (e) for R1 and R3) or else led to a wrong interpretation (roughly equal indicators shown in Subplot (d) for R2). 

\begin{table}[htp!]
\centering
\caption{First 10 resilience indicators obtained by all four tested metrics}
\vspace{0.5cm}
\label{resilience_comparison_table}
\small
\setlength\tabcolsep{2pt}
\begin{tabular}{|l|llll|llll|}
\hline
No. of Cycle &  & NASDAQ &  &  &  & SSE &  &  \\ \cline{2-9} 
 & \multicolumn{1}{l|}{RI} & \multicolumn{1}{l|}{R1} & \multicolumn{1}{l|}{R2} & R3 & \multicolumn{1}{l|}{RI} & \multicolumn{1}{l|}{R1} & \multicolumn{1}{l|}{R2} & R3 \\ \hline
1 & \multicolumn{1}{l|}{1.23} & \multicolumn{1}{l|}{0.00} & \multicolumn{1}{l|}{1.00} & 0.0009 & \multicolumn{1}{l|}{0.51} & \multicolumn{1}{l|}{0.02} & \multicolumn{1}{l|}{0.99} & 0.0023 \\ \hline
2 & \multicolumn{1}{l|}{0.82} & \multicolumn{1}{l|}{-0.07} & \multicolumn{1}{l|}{1.01} & 0.0023 & \multicolumn{1}{l|}{0.04} & \multicolumn{1}{l|}{0.14} & \multicolumn{1}{l|}{0.98} & 0.0010 \\ \hline
3 & \multicolumn{1}{l|}{0.16} & \multicolumn{1}{l|}{0.01} & \multicolumn{1}{l|}{1.00} & 0.0005 & \multicolumn{1}{l|}{2.23} & \multicolumn{1}{l|}{-0.10} & \multicolumn{1}{l|}{1.01} & 0.0018 \\ \hline
4 & \multicolumn{1}{l|}{6.84} & \multicolumn{1}{l|}{0.00} & \multicolumn{1}{l|}{1.00} & 0.0017 & \multicolumn{1}{l|}{0.15} & \multicolumn{1}{l|}{0.97} & \multicolumn{1}{l|}{0.93} & 0.0025 \\ \hline
5 & \multicolumn{1}{l|}{1.88} & \multicolumn{1}{l|}{-0.01} & \multicolumn{1}{l|}{1.00} & 0.0022 & \multicolumn{1}{l|}{18.33} & \multicolumn{1}{l|}{-0.09} & \multicolumn{1}{l|}{1.02} & 0.0017 \\ \hline
6 & \multicolumn{1}{l|}{2.06} & \multicolumn{1}{l|}{0.00} & \multicolumn{1}{l|}{1.00} & 0.0012 & \multicolumn{1}{l|}{0.06} & \multicolumn{1}{l|}{0.09} & \multicolumn{1}{l|}{0.97} & 0.0017 \\ \hline
7 & \multicolumn{1}{l|}{1.11} & \multicolumn{1}{l|}{0.00} & \multicolumn{1}{l|}{1.00} & 0.0024 & \multicolumn{1}{l|}{0.22} & \multicolumn{1}{l|}{0.10} & \multicolumn{1}{l|}{0.98} & 0.0012 \\ \hline
8 & \multicolumn{1}{l|}{2.54} & \multicolumn{1}{l|}{-0.03} & \multicolumn{1}{l|}{1.00} & 0.0011 & \multicolumn{1}{l|}{2.49} & \multicolumn{1}{l|}{-0.01} & \multicolumn{1}{l|}{1.00} & 0.0023 \\ \hline
9 & \multicolumn{1}{l|}{0.76} & \multicolumn{1}{l|}{0.13} & \multicolumn{1}{l|}{0.99} & 0.0021 & \multicolumn{1}{l|}{0.22} & \multicolumn{1}{l|}{0.29} & \multicolumn{1}{l|}{0.96} & 0.0022 \\ \hline
10 & \multicolumn{1}{l|}{0.48} & \multicolumn{1}{l|}{0.06} & \multicolumn{1}{l|}{0.99} & 0.0024 & \multicolumn{1}{l|}{0.18} & \multicolumn{1}{l|}{0.05} & \multicolumn{1}{l|}{0.99} & 0.0008 \\ \hline
\end{tabular}
\end{table}

\begin{figure}[htb!]
	\centering
	\includegraphics[width=1\textwidth]{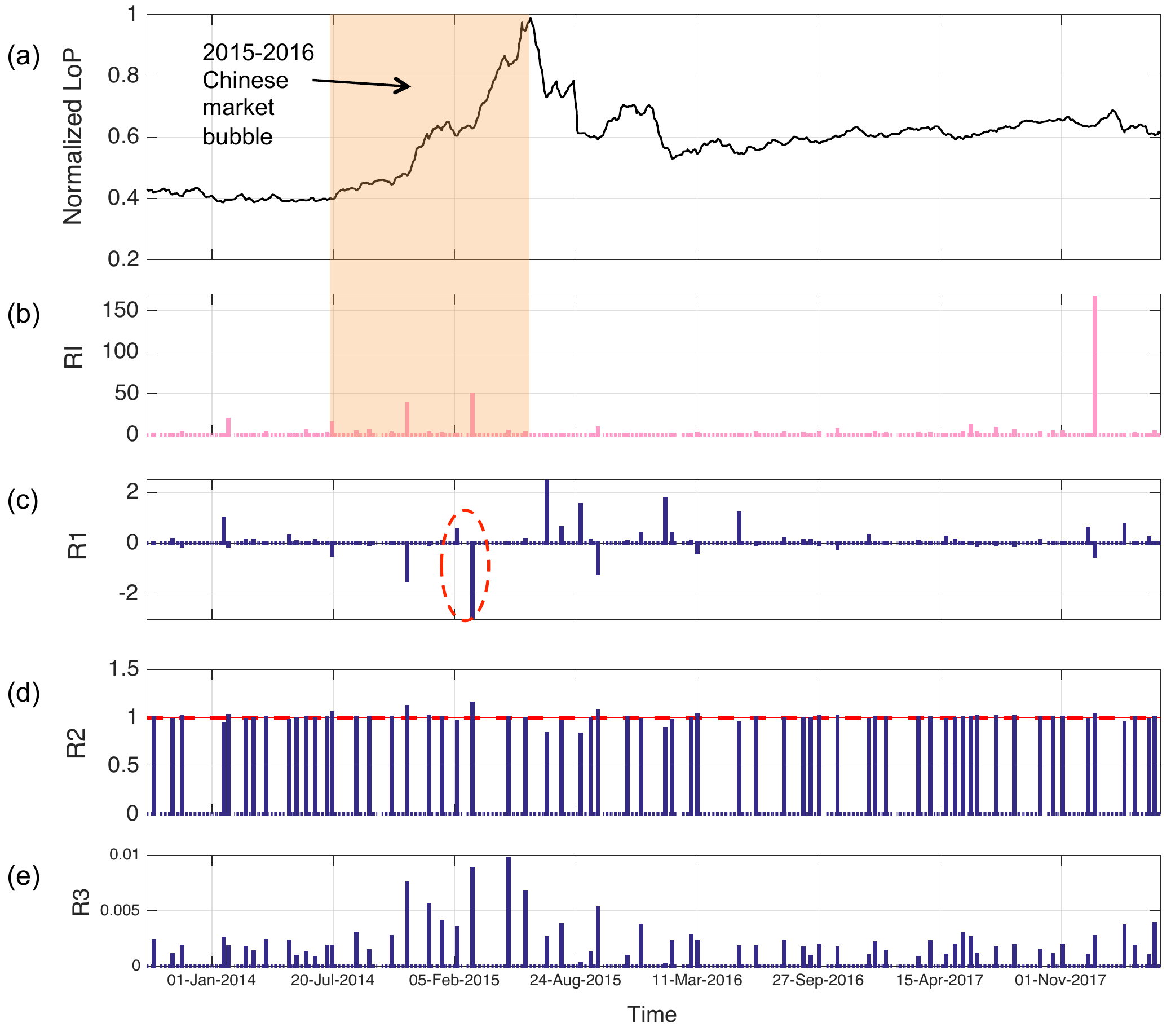}
	\caption{Results from comparisons among four tested metrics, based on the SSE case. (a) normalised LoP, (b) RI, (c) R1, (d) R2, and (e) R3. Obvious defects are highlighted in red. Indicators are plotted at the $t_{event}$ in each cycle.}
	\label{resilience_comparison2} 
\end{figure}

Moreover, the results obtained from R1 and R2 were false. By merely setting the targeted LoP (post-event status) at 100\% of the pre-event level as described, it is obvious that such a consideration was ill-fitted to adaptive performance. The indicators calculated by R1 showed negative values that conflicted with its value range [0, $\infty$] (highlighted in red circle). In addition, the result from R2 contained scores greater than one, which also failed to meet its requirement of [0,1] (highlighted in red line). In subplot~(e), R3 did demonstrate some merits with adaptive recovery as it lacked any fatal mistakes. However, because of its small value range and failure to differentiate strong resilient behaviors from other cycles, its quantification strength was still unpromising. 

Similar observations can be found in the SSE case (Fig~\ref{resilience_comparison2}). There, SSE behaved quite differently than had the NASDAQ during the same period. A major market bubble in the SSE market was also identified according to historical reports~\citep{CNN} during this period. From the quantification results of the proposed metric (subplot (b)), one can see that, prior to the bubble burst, market performance had several strong resilient behaviors as revealed by relatively high resilience scores during that highlighted period. However, after the burst, the following scores dropped dramatically, which correctly indicated a flattened LoP in the overall trend.

\subsection{Dynamics of resilience cycles}
Distribution of the quantified resilience cycles can reveal their dynamic features. Rank-size plots were firstly used to visually identify the empirical distribution of the quantified outcomes, fitted to a theoretical power-law distribution. Secondly, their exceedance probabilities were calculated and plotted in a probability-size graph with log-log scale (log$(P(x_{i}))$vs. log$(x_{i})$) to check the power-law distribution in the upper tail. Finally, the goodness-of-fit tests were conducted by applying the Kolmogorov-Smirnov (K-S) test based on bootstrap re-sampling for 1000 reps to obtain a mean $p$ values. The $p$ values from both cases were used to examine the null hypothesis $H_{0}$: \textit{the data follows a power-law distribution}. In this case, the $H_{0}$ can be rejected with the 0.05 level of significance if $p$ value is $<$ 0.05.

\begin{figure}[]
	\centering
	\includegraphics[width=1\textwidth]{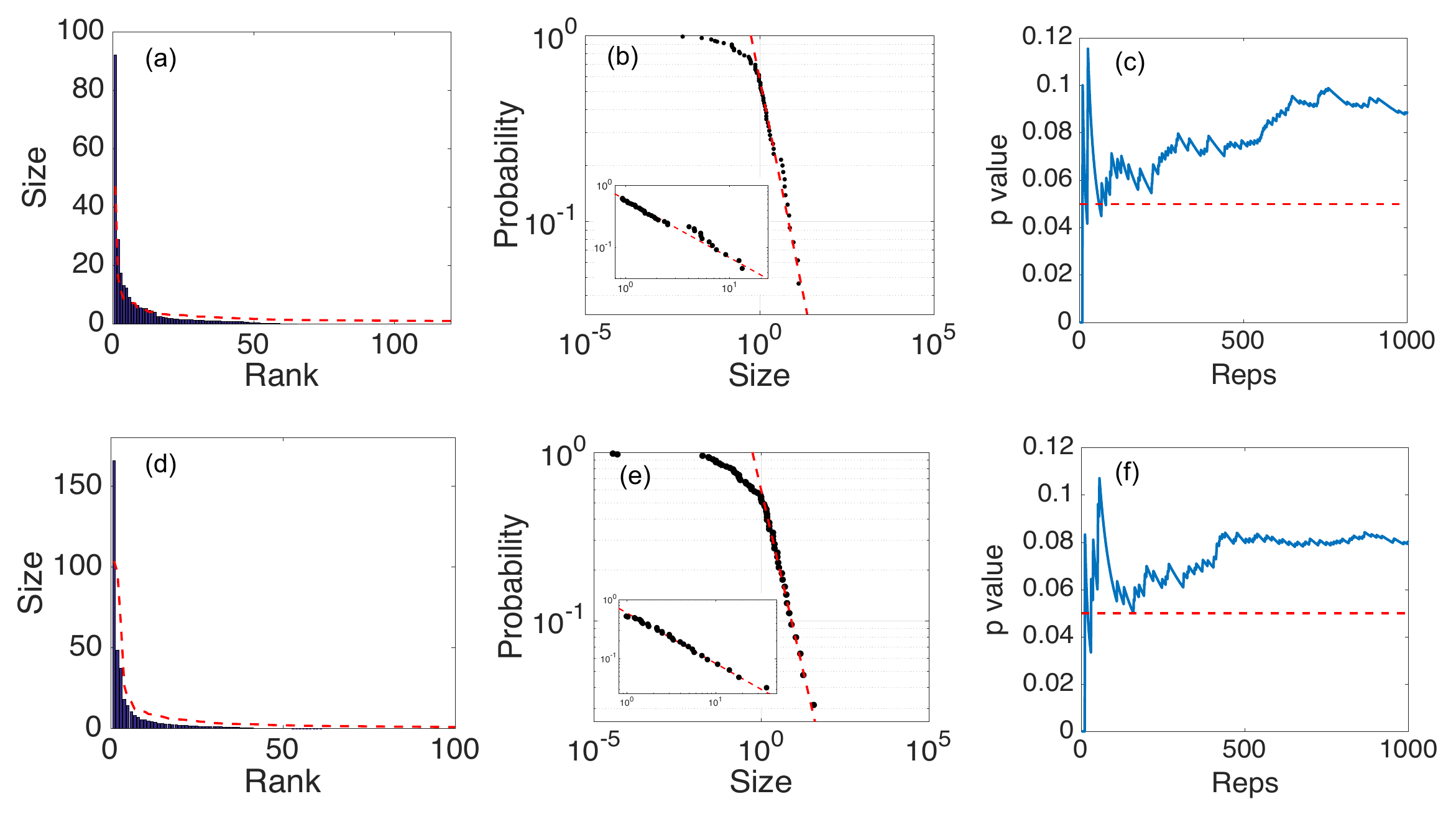}
	\caption{(a) Rank-size plots of the quantified resilience cycles for NASDAQ case. (b) Log-log size-probability plots of the quantified resilience cycles for NASDAQ case, with the inner plot showing specific power-law fit to the tail section. (c) $p$ values of the K-S test with 1000 bootstrap re-sampling test, note that the mean value is above 0.05 significance level. (d-f) corresponding plots for the SSE case.}
	\label{resi_distribution} 
\end{figure}

Figure~\ref{resi_distribution} shows the results of the analysis for both NASDAQ and SSE. Subplot (a) and (d) are rank-size plots of empirical distributions fitted to theoretical power-law distributions, (b) and (e) are log-log plots for fitting the power-law distribution in the upper tail, and (c) and (f) present bootstrap re-sampling results on $p$ values. As seen in the rank-size and probability-size plots, the empirical resilience indicators of consecutive cycles in both cases could be approximated with a power-law distribution in the upper tail. This is especially true in subplot (e), where the model provided an excellent fit in the upper tail of the SSE case and only slight deviance in the NASDAQ case. The mean $p$ values obtained from bootstrap K-S tests granted the findings since both cases had a $p$ value $>$ 0.05, meaning we shall accept the null hypothesis at the significance level of 0.05. For both cases, the power-law distribution and the fat-tail feature indicated a non-trivial dynamics and a strongly stochastic and volatile character in their consecutive resilience cycles.

\subsection{Sensitivity tests}
In this section, the sensitivity test on tolerance thresholds represents a robustness check on the proposed metric, with respect to various levels of investors' psychological capabilities to withstand performance downturns. 

\begin{figure}[htb!]
	\centering
	\includegraphics[width=1\textwidth]{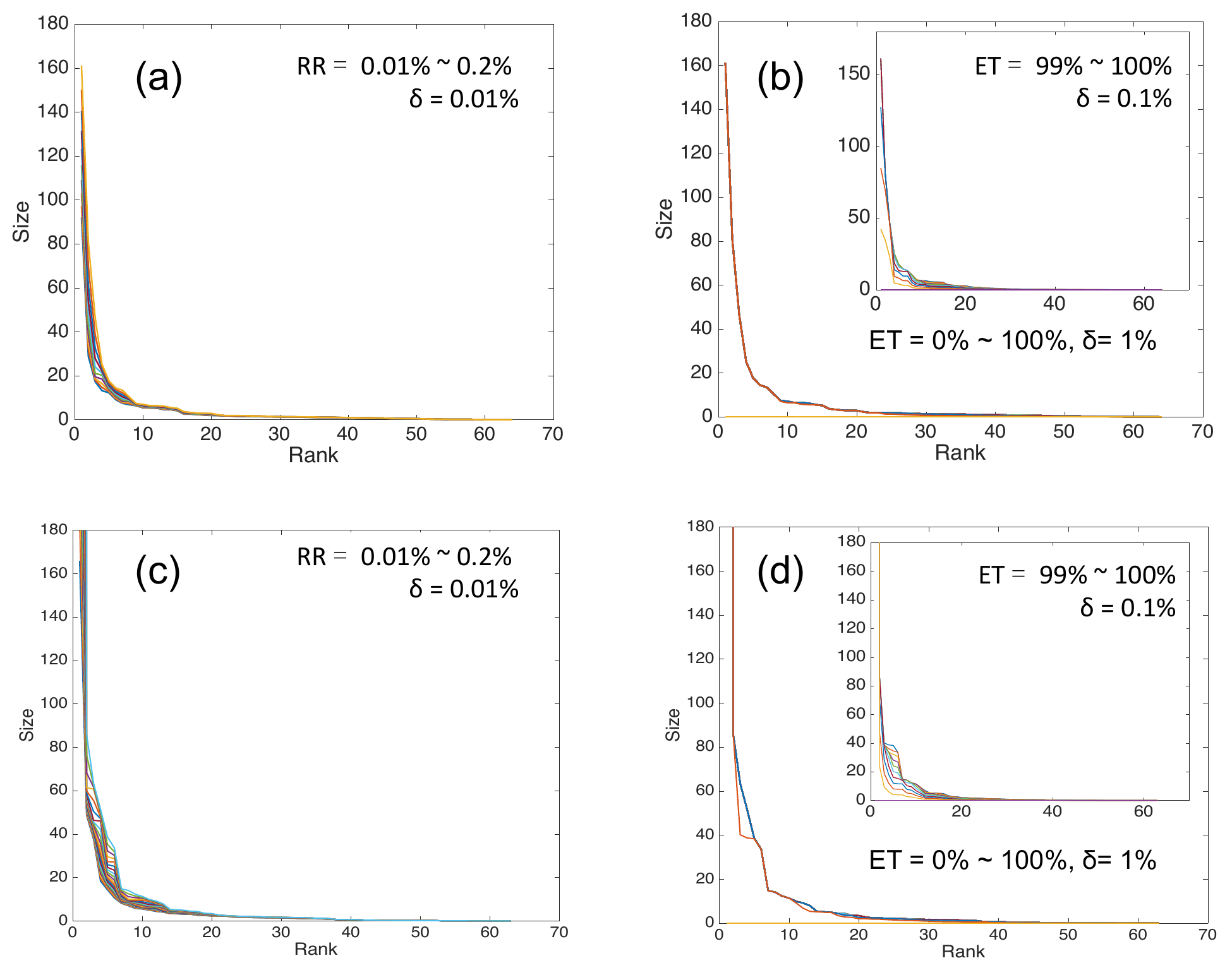}
	\caption{Sensitivity tests for RR and ET. (a) changes in RI values as RR moved from 0.01\% to 0.20\% (0.01\% increments). (b) changes in RI values as ET was adjusted from 0\% to 100\% (1\% increments). Inner plot: micro (0.1\%) steps from 99\% to 100\%. Note that the y-axes of (c) and (d) were limited to 180, same as (a) and (b), for a clear cross-comparison.}
	\label{Sensitivity} 
\end{figure}

Fig~\ref{Sensitivity} shows the sensitivity analysis of resilience indicators as it pertained to these two parameters. For the NASDAQ case, subplot~(a) in this figure shows the analysis of RR varying from 0.01\% to 0.2\% with an incremental step of 0.01\%. Furthermore, subplot~(b) presents results from the analysis of ET, covering the full range (from 0\% to 100\%) at 1\% increments each time with a narrower test range as shown in the inner plot. Subplots (c) and (d) provides the corresponding results for the SSE case. As seen in subplots (a) and (c), the overall performance of the metric was robust to changes in both RR and ET. However, the quantified RI values were relatively sensitive to RR. Cycles with large RI scores were relatively sensitive to RR variations when compared to those with small scores. This implies that perceptions of strong resilient cycles may vary in investors due to different psychological capabilities to withstand downturns.

In contrast, the quantification results tended to be rather insensitive to changes in ET, as shown in subplots (b) and (d). As can be seen, RI values remained relatively stable until they suddenly became zeros once ET reached 100\% (i.e., no capacity was above ET, which meant that $R_{e} = 1$ for all cycles). The term 1 - $R_{e}$ began to vary until ET = 0.99 (inner plot of Fig~\ref{Sensitivity}~(b)). Similar features can also be found in SSE case as shown in subplot(d). These results imply that ET had very limited influence on the final RI values in this case study. Indeed, in most of the resilience cycles, the ET was rarely greater than $min[P(t)]$ because a strongly adaptive and continuously augmented market performance would prevent possible phase transitions. Nevertheless, once the ET was set large enough to allow for a phase transition in most cycles, RI values began to show sensitivity to small variations in ET. In practice, this could imply that large market crashes are rare (such as ``black swan" or ``dragon king" events), but they would significantly alter investors' perception on market resilience when they do occur.

\section{Conclusions}
\label{Conclusion}

This paper proposes a comprehensive function-based resilience metric that takes two fault-tolerance thresholds into account, Robustness Range and Elasticity Threshold, and compare it with three well-established metrics. By taking the time-series performance of two stock markets as empirical studies, this paper studies the applicability of the proposed metric and the dynamics of time-varying resilience in stock market performance while also looking at the effects of tolerance thresholds in investors. The conclusions are summarized as follows:

\begin{itemize}
\item The proposed metric demonstrate satisfactory capability in quantifying time-varying resilience cycles in stock market performance, and it outperforms three well-established metrics in the comparative analysis of applicability.

\item Analysis of distribution shows a strong stochastic character in the dynamics of resilience cycles as quantified by the proposed metric. Here, a power-law distribution is featured in the upper tail based on five years of performance for both tested markets.

\item Sensitivity tests of two tolerance thresholds reveal that the consideration of these two parameters in resilience metric is essential. Even though the proposed metric is, overall, robust to the tuning, large-value resilience cycles are relatively sensitive to RR. In practice, RR represents investors' psychological capability to perceive meaningful downturns. Therefore, the results support that perceptions on market's resilient response could vary among investors.
\end{itemize}

The findings of this paper contribute to our understanding about financial market resilience and explore the effects of tolerance capability on investors' perceptions of the market's resilient response. By taking stock markets as systems, the approach of this paper enriches the resilience assessment toolbox and provides an alternative perspective for interpreting market resilience. The future studies can explore the quantitative assessment of resilient responses in other markets. 

\newpage
\section*{Data access}
The datasets are available to the public on Yahoo finance website.

\section*{Author contribution}
J.T conceived the idea and performed data collection. J.T analyzed the data and wrote the paper with supervisions from H.R.H. All authors gave final approval to the manuscript.

\section*{Competing interest}
The authors declare no conflicts of interest.

\section*{Funding}
This research was funded by ETH Zurich and Singapore's National Research Foundation (FI 370074011).

\section*{Abbreviations}
ET: Elasticity threshold.
LoP: Level of performance.
MCHP: Multi-cycle hybrid performance.
RI: Resilience indicator.
RR: Robustness range.

\section*{Acknowledgement}
This research was conducted at the Future Resilient Systems at the Singapore-ETH Centre, which was established collaboratively between ETH Zurich and Singapore's National Research Foundation (FI 370074011) under its Campus for Research Excellence and Technological Enterprise programme. We would like to thank Dr. Xing Zhang from ETH Zurich for his discussion on the paper content.

\newpage
\section*{Appendix A: Implementation procedure for the proposed metric}
\begin{figure}[htb!]
	\centering
	\includegraphics[width=1\textwidth]{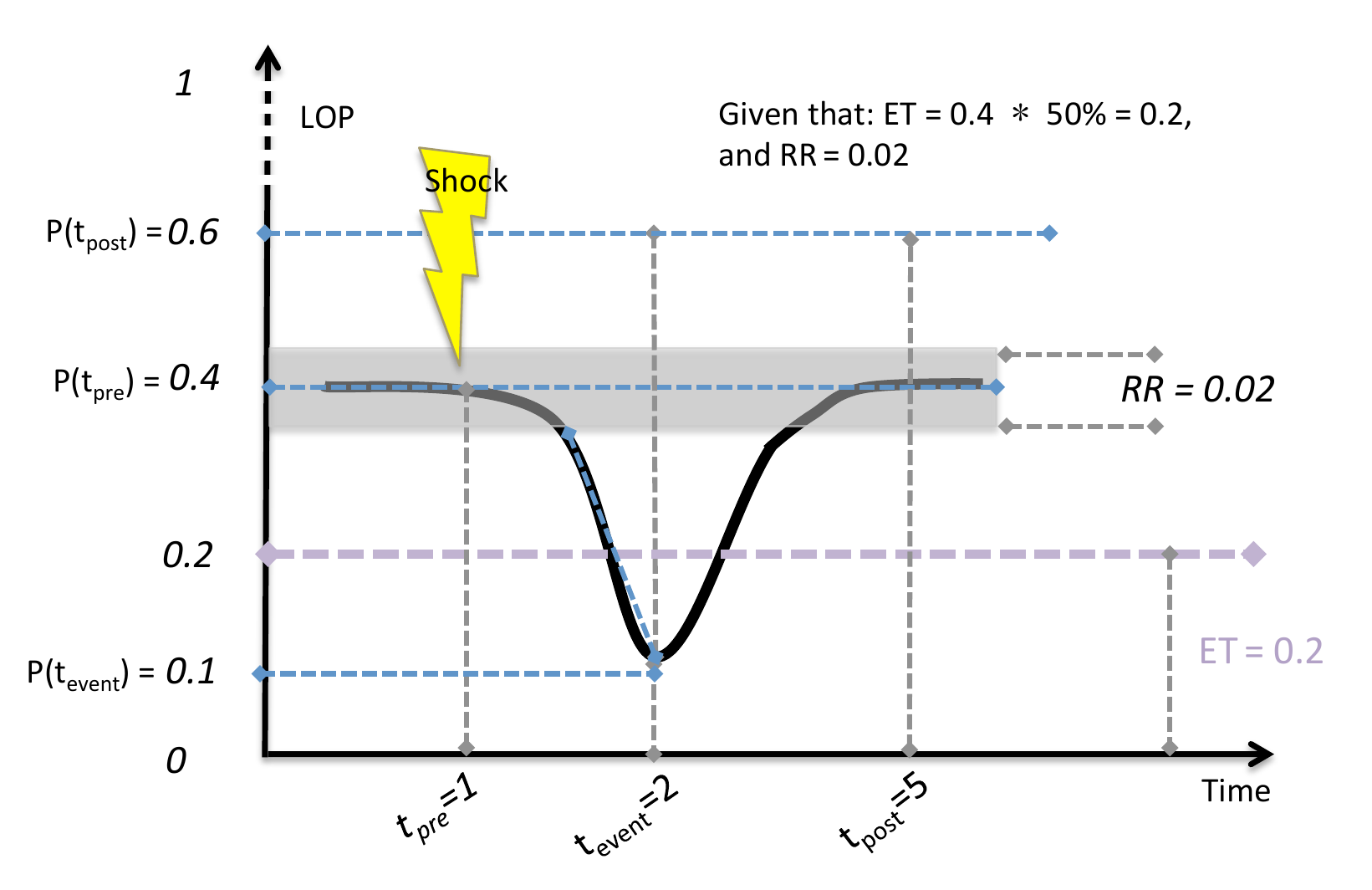}
	\caption{A typical down-up cycle associated with the ``Leveled recovery" scenario (values not to scale). (Color is not needed)}
	\label{metric_calculation} 
\end{figure}

For clarification, an illustrative example of how this proposed metric can be implemented in each identified cycle is provided. Fig~\ref{metric_calculation} presents a typical drawdown-and-drawup cycle under the ``Leveled recovery" scenario. Given that the $P_{ET}$ and $P_{RR}$ are 50\% and 1\%, respectively.

\begin{enumerate}
\item Equation (3.1-3.2): \textbf{ET} is calculated as $P_{ET} \times P(t_{pre})$ = $0.4 \times 50\% = 0.2$. Also, \textbf{RR} = $\pm 1\% = 0.02$
\item Equation (3.3): \textbf{Resistance} $R_{m} = 0.1 + 0.02 = 0.12$
\item Equation (3.4): \textbf{Re-stabilization} Because ET $> min[P(t)]$, $R_{e} = \frac{0.2 - 0.1}{0.2} = 0.5$
\item Equation (3.5-3.6): \textbf{Rebuilding} $S_{f} = \frac{0.4 - 0.1}{2-1} = 0.3$ and $S_{r} = \frac{0.4 - 0.1}{5 - 2} = 0.1$. Thus, $R_{d} = \frac{0.1}{0.3} = 0.333$.
\item Equation (3.7): \textbf{Reconfiguration} $R_{s} = \frac{0.3}{0.3} = 1$ for ``Just recovery" scenario.
\item Thus, Equation (3.8): $RI = 0.12 \times (1- 0.5) \times 0.333 \times 1 = 0.020$ for this example
\end{enumerate}

 \singlespacing

\cleardoublepage
\bibliographystyle{apa} 
\bibliography{/Users/junqing/Desktop/1_Papers_data_Codes/Working_papers/Paper_1_ResiModel_finished/Revised_version4/Final} 

\begin{thebibliography}{}

\bibitem[\protect\astroncite{Bhamra et~al.}{2011}]{bhamra2011resilience}
Bhamra, R., Dani, S., and Burnard, K. (2011).
\newblock Resilience: The concept, a literature review and future directions.
\newblock {\em International Journal of Production Research},
  49(18):5375--5393.

\bibitem[\protect\astroncite{Biringer et~al.}{2016}]{biringer2016critical}
Biringer, B., Vugrin, E., and Warren, D. (2016).
\newblock {\em Critical infrastructure system security and resiliency}.
\newblock CRC press.

\bibitem[\protect\astroncite{Bookstaber et~al.}{2016}]{bookstaber2016toward}
Bookstaber, R., Foley, M.~D., and Tivnan, B.~F. (2016).
\newblock Toward an understanding of market resilience: market liquidity and
  heterogeneity in the investor decision cycle.
\newblock {\em Journal of Economic Interaction and Coordination},
  11(2):205--227. (doi:10.1007/s11403--015--0162--8).

\bibitem[\protect\astroncite{Bruneau et~al.}{2003}]{bruneau2003framework}
Bruneau, M., Chang, S.~E., Eguchi, R.~T., Lee, G.~C., O'Rourke, T.~D.,
  Reinhorn, A.~M., Shinozuka, M., Tierney, K., Wallace, W.~A., and von
  Winterfeldt, D. (2003).
\newblock A framework to quantitatively assess and enhance the seismic
  resilience of communities.
\newblock {\em Earthquake spectra}, 19(4):733--752.

\bibitem[\protect\astroncite{Bruneau and Reinhorn}{2006}]{bruneau2006overview}
Bruneau, M. and Reinhorn, A. (2006).
\newblock Overview of the resilience concept.
\newblock In {\em Proceedings of the 8th US National Conference on Earthquake
  Engineering. USA.}, pages 18--22.

\bibitem[\protect\astroncite{Cere et~al.}{2017}]{Cere2017}
Cere, G., Rezgui, Y., and Zhao, W. (2017).
\newblock Critical review of existing built environment resilience frameworks:
  Directions for future research.
\newblock {\em International Journal of Disaster Risk Reduction}, 25:173--189.

\bibitem[\protect\astroncite{Cimellaro et~al.}{2010}]{cimellaro2010framework}
Cimellaro, G.~P., Reinhorn, A.~M., and Bruneau, M. (2010).
\newblock Framework for analytical quantification of disaster resilience.
\newblock {\em Engineering Structures}, 32(11):3639--3649.

\bibitem[\protect\astroncite{Couzin-Frankel}{2018}]{couzin2018roots}
Couzin-Frankel, J. (2018).
\newblock The roots of resilience.
\newblock {\em Science (New York, NY)}, 359(6379):970--971.

\bibitem[\protect\astroncite{Drakos}{2011}]{drakos2011behavioral}
Drakos, K. (2011).
\newblock Behavioral channels in the cross-market diffusion of major terrorism
  shocks.
\newblock {\em Risk Analysis: An International Journal}, 31(1):143--159.

\bibitem[\protect\astroncite{Duval et~al.}{2007}]{duval2007structural}
Duval, R., Elmeskov, J., and Vogel, L. (2007).
\newblock Structural policies and economic resilience to shocks.
\newblock {\em Organisation for Economic Co-operation and Development
  (OECD-SSRN)}.

\bibitem[\protect\astroncite{Erragragui et~al.}{2018}]{erragragui2018does}
Erragragui, E., Hassan, M.~K., Peillex, J., and Khan, A. N.~F. (2018).
\newblock Does ethics improve stock market resilience in times of instability?
\newblock {\em Economic Systems}.

\bibitem[\protect\astroncite{Farmer et~al.}{2012}]{farmer2012complex}
Farmer, J.~D., Gallegati, M., Hommes, C., Kirman, A., Ormerod, P., Cincotti,
  S., Sanchez, A., and Helbing, D. (2012).
\newblock A complex systems approach to constructing better models for managing
  financial markets and the economy.
\newblock {\em The European Physical Journal Special Topics}, 214(1):295--324.

\bibitem[\protect\astroncite{Filimonov and Sornette}{2015}]{filimonov2015power}
Filimonov, V. and Sornette, D. (2015).
\newblock Power law scaling and ``dragon-kings'' in distributions of intraday
  financial drawdowns.
\newblock {\em Chaos, Solitons \& Fractals}, 74:27--45.

\bibitem[\protect\astroncite{Fisher}{2015}]{fisher2015disaster}
Fisher, L. (2015).
\newblock Disaster responses: More than 70 ways to show resilience.
\newblock {\em Nature}, 518(7537):35.

\bibitem[\protect\astroncite{Francis and Bekera}{2014}]{francis2014metric}
Francis, R. and Bekera, B. (2014).
\newblock A metric and frameworks for resilience analysis of engineered and
  infrastructure systems.
\newblock {\em Reliability Engineering \& System Safety}, 121:90--103.

\bibitem[\protect\astroncite{Gal{\'\i} and Gambetti}{2015}]{gali2015effects}
Gal{\'\i}, J. and Gambetti, L. (2015).
\newblock The effects of monetary policy on stock market bubbles: Some
  evidence.
\newblock {\em American Economic Journal: Macroeconomics}, 7(1):233--57.

\bibitem[\protect\astroncite{Gonz{\'a}lez et~al.}{1997}]{gonzalez1997adaptive}
Gonz{\'a}lez, O., Shrikumar, H., Stankovic, J.~A., and Ramamritham, K. (1997).
\newblock Adaptive fault tolerance and graceful degradation under dynamic hard
  real-time scheduling.
\newblock In {\em Real-Time Systems Symposium, 1997. Proceedings., The 18th
  IEEE}, pages 79--89.

\bibitem[\protect\astroncite{Haimes}{2009}]{haimes2009complex}
Haimes, Y.~Y. (2009).
\newblock On the complex definition of risk: A systems-based approach.
\newblock {\em Risk Analysis}, 29(12):1647--1654.

\bibitem[\protect\astroncite{Hashimoto et~al.}{1982}]{hashimoto1982reliability}
Hashimoto, T., Stedinger, J.~R., and Loucks, D.~P. (1982).
\newblock Reliability, resiliency, and vulnerability criteria for water
  resource system performance evaluation.
\newblock {\em Water Resources Research}, 18(1):14--20.

\bibitem[\protect\astroncite{Heinimann}{2016}]{Heinimann2016}
Heinimann, H.~R. (2016).
\newblock A generic framework for resilience assessment.
\newblock {\em In Resource Guide on Resilience. IRGC, Editor. EPFL
  International Risk Governance Center: Lausanne.}

\bibitem[\protect\astroncite{Heinimann and Hatfield}{2017}]{HeinimannHatfield}
Heinimann, H.~R. and Hatfield, K. (2017).
\newblock Infrastructure resilience assessment, management and governance:
  State and perspectives.
\newblock {\em NATO Science for Peace and Security Series C: Environmental
  Security}, pages 147--187.

\bibitem[\protect\astroncite{Hill et~al.}{2012}]{hill2012economic}
Hill, E., Clair, T.~S., Wial, H., Wolman, H., Atkins, P., Blumenthal, P.,
  Ficenec, S., and Friedhoff, A. (2012).
\newblock Economic shocks and regional economic resilience.
\newblock In {\em Urban and Regional Policy and Its Effects: Building Resilient
  Regions}, pages 193--274. Brookings Institution Press.

\bibitem[\protect\astroncite{Hosseini et~al.}{2016}]{hosseini2016review}
Hosseini, S., Barker, K., and Ramirez-Marquez, J.~E. (2016).
\newblock A review of definitions and measures of system resilience.
\newblock {\em Reliability Engineering \& System Safety}, 145:47--61.

\bibitem[\protect\astroncite{Johansen and Sornette}{1998}]{johansen1998stock}
Johansen, A. and Sornette, D. (1998).
\newblock Stock market crashes are outliers.
\newblock {\em The European Physical Journal B-Condensed Matter and Complex
  Systems}, 1(2):141--143.

\bibitem[\protect\astroncite{Johansen and Sornette}{2000}]{johansen2000large}
Johansen, A. and Sornette, D. (2000).
\newblock Large stock market price drawdowns are outliers.
\newblock {\em Available at SSRN 244563.}, pages 1--45.

\bibitem[\protect\astroncite{Johansen et~al.}{2010}]{johansen2010shocks}
Johansen, A., Sornette, D., et~al. (2010).
\newblock Shocks, crashes and bubbles in financial markets.
\newblock {\em Brussels Economic Review}, 53(2):201--253.

\bibitem[\protect\astroncite{Joo et~al.}{2017}]{joo2017long}
Joo, K., Shim, H.~S., and Sul, W. (2017).
\newblock The long-run effect of environmental issues on stock market
  performance: Evidence from the us stock market.
\newblock {\em International Business Journal}, 28(3):101--133.
  (doi:10.14365/ibj.2017.28.3.4).

\bibitem[\protect\astroncite{Kacperski and Holyst}{1996}]{kacperski1996phase}
Kacperski, K. and Holyst, J.~A. (1996).
\newblock Phase transitions and hysteresis in a cellular automata-based model
  of opinion formation.
\newblock {\em Journal of statistical physics}, 84(1-2):169--189.
  (doi:10.1007/bf02179581).

\bibitem[\protect\astroncite{Kaizoji}{2000}]{kaizoji2000speculative}
Kaizoji, T. (2000).
\newblock Speculative bubbles and crashes in stock markets: An
  interacting-agent model of speculative activity.
\newblock {\em Physica A: Statistical Mechanics and its Applications},
  287(3-4):493--506.

\bibitem[\protect\astroncite{Kau{\^e} Dal'Maso~Peron
  et~al.}{2012}]{kaue2012structure}
Kau{\^e} Dal'Maso~Peron, T., da~Fontoura~Costa, L., and Rodrigues, F.~A.
  (2012).
\newblock The structure and resilience of financial market networks.
\newblock {\em Chaos: An Interdisciplinary Journal of Nonlinear Science},
  22(1):013117.

\bibitem[\protect\astroncite{Leal and Napoletano}{2017}]{leal2017market}
Leal, S.~J. and Napoletano, M. (2017).
\newblock Market stability vs. market resilience: Regulatory policies
  experiments in an agent-based model with low-and high-frequency trading.
\newblock {\em Journal of Economic Behavior \& Organization}, page In press.
  (doi:10.1016/j.jebo.2017.04.013).

\bibitem[\protect\astroncite{Lehkonen and
  Heimonen}{2015}]{lehkonen2015democracy}
Lehkonen, H. and Heimonen, K. (2015).
\newblock Democracy, political risks and stock market performance.
\newblock {\em Journal of International Money and Finance}, 59:77--99.
  (doi:10.1016/j.jimonfin.2015.06.002).

\bibitem[\protect\astroncite{Lilienfeld-toal and
  Ruenzi}{2014}]{lilienfeld2014ceo}
Lilienfeld-toal, U.~V. and Ruenzi, S. (2014).
\newblock Ceo ownership, stock market performance, and managerial discretion.
\newblock {\em The Journal of Finance}, 69(3):1013--1050.
  (doi:10.1016/j.jimonfin.2015.06.002).

\bibitem[\protect\astroncite{Linkov et~al.}{2013}]{linkov2013measurable}
Linkov, I., Eisenberg, D.~A., Bates, M.~E., Chang, D., Convertino, M., Allen,
  J.~H., Flynn, S.~E., and Seager, T.~P. (2013).
\newblock Measurable resilience for actionable policy.
\newblock {\em Environmental Science \& Technology}, 47(18):10108--10110.

\bibitem[\protect\astroncite{Martin}{2011}]{martin2011regional}
Martin, R. (2011).
\newblock Regional economic resilience, hysteresis and recessionary shocks.
\newblock {\em Journal of Economic Geography}, 12(1):1--32.

\bibitem[\protect\astroncite{Martin-Breen and
  Anderies}{2011}]{martin2011resilience}
Martin-Breen, P. and Anderies, J.~M. (2011).
\newblock Resilience: A literature review.
\newblock Technical report, Brighton: Institute of Development Studies,
  Brighton, UK.

\bibitem[\protect\astroncite{Mathworks}{2016}]{CurveFitting}
Mathworks (2016).
\newblock Filtering and smoothing data.
\newblock Technical report, Matlab R2018a Documentations. [Available at]
  http://www.mathworks.com/help/curvefit/smoothing-data.html.

\bibitem[\protect\astroncite{Nan and Sansavini}{2017}]{nan2017quantitative}
Nan, C. and Sansavini, G. (2017).
\newblock A quantitative method for assessing resilience of interdependent
  infrastructures.
\newblock {\em Reliability Engineering \& System Safety}, 157:35--53.

\bibitem[\protect\astroncite{Nan et~al.}{2014}]{nan2014building}
Nan, C., Sansivini, G., and Kr{\"o}ger, W. (2014).
\newblock Building an integrated metric for quantifying the resilience of
  interdependent infrastructure systems.
\newblock In {\em International Conference on Critical Information
  Infrastructures Security}, pages 159--171. Springer.

\bibitem[\protect\astroncite{Ouyang and
  Due{\~n}as-Osorio}{2012}]{ouyang2012time}
Ouyang, M. and Due{\~n}as-Osorio, L. (2012).
\newblock Time-dependent resilience assessment and improvement of urban
  infrastructure systems.
\newblock {\em Chaos: An Interdisciplinary Journal of Nonlinear Science},
  22(3):033122.

\bibitem[\protect\astroncite{Pereira et~al.}{2018}]{pereira2018trump}
Pereira, E. J. d. A.~L., da~Silva, M.~F., da~Cunha~Lima, I., and Pereira, H.
  (2018).
\newblock Trump's effect on stock markets: A multiscale approach.
\newblock {\em Physica A: Statistical Mechanics and its Applications},
  512:241--247.

\bibitem[\protect\astroncite{Riley and Yan}{2015}]{CNN}
Riley, C. and Yan, S. (2015).
\newblock China's stock market crash...in 2 minutes.
\newblock Technical report, CNN Money. [Available at]
  http://money.cnn.com/2015/07/09/investing/china-crash-in-two-minutes/index.html.

\bibitem[\protect\astroncite{Rose}{2004}]{rose2004defining}
Rose, A. (2004).
\newblock Defining and measuring economic resilience to disasters.
\newblock {\em Disaster Prevention and Management: An International Journal},
  13(4):307--314.

\bibitem[\protect\astroncite{Rose}{2007}]{rose2007economic}
Rose, A. (2007).
\newblock Economic resilience to natural and man-made disasters:
  Multidisciplinary origins and contextual dimensions.
\newblock {\em Environmental Hazards}, 7(4):383--398.

\bibitem[\protect\astroncite{Rus et~al.}{2018}]{rus2018resilience}
Rus, K., Kilar, V., and Koren, D. (2018).
\newblock Resilience assessment of complex urban systems to natural disasters:
  A new literature review.
\newblock {\em International Journal of Disaster Risk Reduction}, pages
  311--330.

\bibitem[\protect\astroncite{Simmie and Martin}{2010}]{simmie2010economic}
Simmie, J. and Martin, R. (2010).
\newblock The economic resilience of regions: towards an evolutionary approach.
\newblock {\em Cambridge Journal of Regions, Economy and Society}, 3(1):27--43.

\bibitem[\protect\astroncite{Sornette}{2004}]{sornette2004complex}
Sornette, D. (2004).
\newblock A complex system view of why stock markets crash.
\newblock {\em New Thesis}, 1(1):5--18.

\bibitem[\protect\astroncite{Tang and Heinimann}{2018}]{tang2018resilience}
Tang, J. and Heinimann, H.~R. (2018).
\newblock A resilience-oriented approach for quantitatively assessing recurrent
  spatial-temporal congestion on urban roads.
\newblock {\em PLoS One}, 13(1):e0190616.

\bibitem[\protect\astroncite{Tang et~al.}{2017}]{tang2017modeling}
Tang, J., Khoja, L., and Heinimann, H.~R. (2017).
\newblock Modeling stock survivability resilience in signed temporal networks:
  A study from london stock exchange.
\newblock In {\em International Workshop on Complex Networks and their
  Applications}, pages 1041--1052. Springer.

\bibitem[\protect\astroncite{Tang et~al.}{2018}]{tang2018characterisation}
Tang, J., Khoja, L., and Heinimann, H.~R. (2018).
\newblock Characterisation of survivability resilience with dynamic stock
  interdependence in financial networks.
\newblock {\em Applied Network Science}, 3(1):23.

\bibitem[\protect\astroncite{Tierney and
  Bruneau}{2007}]{tierney2007conceptualizing}
Tierney, K. and Bruneau, M. (2007).
\newblock Conceptualizing and measuring resilience: A key to disaster loss
  reduction.
\newblock {\em TR news}, 250(250).

\bibitem[\protect\astroncite{Wein and Rose}{2011}]{wein2011economic}
Wein, A. and Rose, A. (2011).
\newblock Economic resilience lessons from the shakeout earthquake scenario.
\newblock {\em Earthquake Spectra}, 27(2):559--573.

\bibitem[\protect\astroncite{Wili{\'n}ski
  et~al.}{2013}]{wilinski2013structural}
Wili{\'n}ski, M., Sienkiewicz, A., Gubiec, T., Kutner, R., and Struzik, Z.
  (2013).
\newblock Structural and topological phase transitions on the german stock
  exchange.
\newblock {\em Physica A: Statistical Mechanics and its Applications},
  392(23):5963--5973.

\bibitem[\protect\astroncite{Yahoo}{2018}]{YahooFinance}
Yahoo (2018).
\newblock Historical data of nasdaq composite. [available at]
  https://www.google.com/finance/historical?q=indexnasdaq

\bibitem[\protect\astroncite{Zhao et~al.}{2018}]{zhao2018stock}
Zhao, L., Wang, G.-J., Wang, M., Bao, W., Li, W., and Stanley, H.~E. (2018).
\newblock Stock market as temporal network.
\newblock {\em Physica A: Statistical Mechanics and its Applications},
  506:1104--1112.

\end{thebibliography}

\end{document}